\documentclass[12pt]{iopart}
\usepackage{iopams}
\usepackage{epsfig}
\usepackage{textcomp}
\usepackage{citesort}
\begin{document}

\title[Information sharing promotes prosocial behaviour]{Information sharing promotes prosocial behaviour}

\author{Attila Szolnoki$^1$ and Matja{\v z} Perc$^2$}
\address{$^1$Institute of Technical Physics and Materials Science, Research Centre for Natural Sciences, Hungarian Academy of Sciences, P.O. Box 49, H-1525 Budapest, Hungary\\
$^2$Department of Physics, Faculty of Natural Sciences and Mathematics, University of Maribor, Koro{\v s}ka  cesta 160, SI-2000 Maribor, Slovenia}
\ead{szolnoki.attila@ttk.mta.hu, matjaz.perc@uni-mb.si}

\begin{abstract}
More often than not, bad decisions are bad regardless of where and when they are made. Information sharing might thus be utilized to mitigate them. Here we show that sharing the information about strategy choice between players residing on two different networks reinforces the evolution of cooperation. In evolutionary games the strategy reflects the action of each individual that warrants the highest utility in a competitive setting. We therefore assume that identical strategies on the two networks reinforce themselves by lessening their propensity to change. Besides network reciprocity working in favour of cooperation on each individual network, we observe the spontaneous emerge of correlated behaviour between the two networks, which further deters defection. If information is shared not just between individuals but also between groups, the positive effect is even stronger, and this despite the fact that information sharing is implemented without any assumptions with regards to content.
\end{abstract}

\pacs{87.23.Ge, 89.75.Fb, 89.65.-s}
\maketitle

\section{Introduction}
\label{intro}
We create an enormous amount of information on a daily basis. According to Google's Executive Chairman Eric Schmidt, every two days as much as we did from the beginning of time up to 2003. It is the availability of this information that fuels data-driven research efforts aimed at understanding the geographical patterns of mobility \cite{gonzales_n08,song_s10} and scientific production \cite{pan_sr12}, the limits of predictability of sport performance \cite{radicchi_pone12} and market change \cite{preis_sr12}, the spread of infectious diseases \cite{balcan_pnas09,meloni_pnas09} and malware \cite{hu_pnas09,wang_p_s09}, as well as the dynamics of online popularity \cite{ratkiewicz_prl10}, social movements \cite{borge-holthoefer_pone11} and political campaigns \cite{bond_n12}, to name but a few examples. Indeed, making information available contributes to discoveries across the whole spectrum of social and natural sciences \cite{lazer_s09}. Withholding information or failing to integrate it properly into the bigger picture, on the other hand, can have many unwanted and unintended consequences \cite{ball_12}.

Research in the realm of network science has recently highlighted that seemingly irrelevant changes in one network can have catastrophic and very much unexpected consequences in another network \cite{buldyrev_n10,zhou_d_pre12}. The key to understanding these phenomena is network interdependence \cite{buldyrev_n10,parshani_prl10,huang_xq_pre11,gao_jx_np12,gomez_prl13}, and in particular knowing how information available in one network might affect behaviour in another network. Figure~\ref{scheme} depicts the red node on the upper network wanting to enforce its state (strategy) on the blue node. Based on the information stemming from the bottom network, however, the blue node might be more reluctant to the change than in the absence of that information. This example can be made more concrete in the realm of evolutionary games on networks \cite{szabo_pr07,schuster_jbp08,roca_plr09,perc_jrsi13}, where the players compete for space by choosing to cooperate or defect based on their success in maximizing their utility. The prisoner's dilemma \cite{maynard_82,axelrod_84}, for example, promises a defector the temptation $T>1$ when facing a cooperator, while two cooperators receive only the reward $R=1$ each. Each individual is therefore tempted to defect. Yet such antisocial behaviour can lead to the tragedy of the commons \cite{hardin_g_s68}. On a single network \cite{kim_bj_pre02,zimmermann_pre04,santos_prl05,santos_pnas06,ohtsuki_n06,gomez-gardenes_prl07,floria_pre09,roca_pre09,antonioni_pone11,du_f_dga11,portillo_pre12,allen_jtb12}, various forms of reciprocity \cite{nowak_s06,pacheco_ploscb09,perc_bs10,pacheco_jtb08}, most notably network reciprocity \cite{nowak_n92b}, promote the evolution of cooperation. If the networks are more than one and interdependent \cite{vukov_pre05,wang_j_epl11,gomez-gardenes_pre12,gomez-gardenes_srep12,wang_z_epl12}, new phenomena may emerge that additionally favour prosocial behaviour. The enhanced resilience of cooperation can be due to a non-trivial organization of cooperation across different network layers.

Unlike in previous works, we do not consider players in one network actually affecting the utility of players in the other network. Instead, solely the information about the strategies adopted in the other network is transmitted, and this affects the propensity of players to change their strategy. While the probability to change strategy is still determined by the difference in utility between the two neighbouring players in a given network, e.g., the red and blue node in the upper network of Fig.~\ref{scheme}, the information about which strategies are adopted in the other network is used to scale this probability. In particular, if the corresponding players in the other network adopt the same strategy as the player targeted with a new strategy, as is the case in Fig.~\ref{scheme}, the probability to change strategy is decreased proportionally with the number of such players locally present in the other network. Accordingly, even though the utility stemming from the prisoner's dilemma game of the red player in Fig.~\ref{scheme} may be much larger than that of the neighbouring blue player, the probability to change strategy to red will be low because all the corresponding players in the bottom network are adopting the blue strategy.

Many real life examples can be given to support such a procedure, not least insisting on regulatory policies in one country based on the success of the same policies in another country. Importantly, we make no assumptions with regards to the content of the information that is transmitted. Defectors are just as free to transmit their recommendation to the other network as cooperators. In this sense, the information transfer is strategy neutral, and the information itself is neither filtered nor evaluated based on perceived importance or potential impact. These are important assumptions given that in general it is impossible to know in advance which information will be useful and how it will affect the receiver. We only assume that the information about the strategies \textit{is} shared, and as we will show, this alone is sufficient for prosocial behaviour to be promoted.

The remainder of this paper is organized as follows. In the next section, we describe the model and the considered evolutionary games. Section 3 is devoted to the presentation of results, whereas lastly, we summarize the main conclusions and discuss their potential implications.

\section{Model}
\label{model}

\begin{figure}
\centerline{\epsfig{file=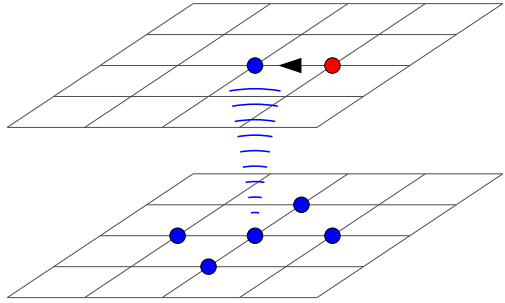,width=7cm}}
\caption{\label{scheme} Information sharing between networks affects strategy transfer between neighbouring players. The red player in the upper network tries to transfer its strategy to the blue player. The blue player receives information from the corresponding players in the bottom network that they all adopt the blue strategy. Because of this, the blue player in the upper network is reluctant to change strategy to red, despite the fact that the red player might have a higher utility.}
\end{figure}

The prisoner's dilemma game on both networks is staged on a $L \times L$ square lattice with periodic boundary conditions, where each player is connected with its $k=4$ nearest neighbors, as depicted in Fig.~\ref{scheme}. Initially each player on site $x$ in network A (up) and on site $x^\prime$ in network B (bottom) is designated either as a cooperator ($C$) or defector ($D$) with equal probability. The accumulation of payoffs $\pi_x$ and $\pi_{x^\prime}$ on both networks follows the same procedure. Namely, each player plays the prisoner's dilemma game with its four nearest neighbours, whereby mutual cooperation yields the reward $R=1$, mutual defection and cooperation while the neighbour defects yields zero, and defecting while the neighbour cooperates yields the temptation $T>1$. The same parametrization has been adopted several times before \cite{nowak_n92b}, and it is accepted that it captures all relevant aspects of the prisoner's dilemma game. More importantly, it also enables a relevant comparison with the many preceding studies.

Following the determination of payoffs, which we here consider to be fully representative for the utility, strategy imitation is possible only between nearest neighbours on any given lattice, but never between players residing on different networks. Accordingly, on network A player $x$ can adopt the strategy $s_{y}$ of one of its randomly chosen nearest neighbours $y$ with a probability determined by the Fermi function \cite{szabo_pr07}
\begin{equation}
W(s_x \leftarrow s_y) = w_x  \frac{1}{1+\exp[(\pi_{s_x}-\pi_{s_y})/K]}\,\,.
\end{equation}
Here the scaling factor $w_x$ of player $x$ depends on the strategies of related players from the other network, and as such it is the key quantity that takes into account information sharing.

The simplest option is to consider only the strategy of the directly corresponding player $x^{\prime}$ on the other network, and assume that $w_x$ is minimal if $s_x = s_{x^{\prime}}$ and maximal in the opposite case. To avoid frozen states we use $w_{\min}=0.1$ as the minimal scaling factor, while the maximal is $w_{\max}=1$. We will refer to this model as model $S$, because only a single player in the other network is taken as reference for determining $w_x$. An extension of the simplest model is if not only $x^\prime$ but also its neighbors determine $w_x$, as schematically depicted in Fig.~\ref{scheme}. In this case $w_x$ changes linearly between $w_{\min}$ and $w_{\max}$ in accordance with
\begin{equation}
w_x = 1 - (w_{\max}-w_{\min})\frac{N_x}{G} \,,
\end{equation}
where $G=k+1$ is the size of the considered group in the other network and $N_x$ is the number of players in that group that adopt the same strategy as player $x$. We will refer to this model as model $G$, because a whole group in the other network is taken as reference for determining $w_x$.

Since the prisoner's dilemma game is governed by pairwise interactions, we test the robustness of presented results by employing also the public goods game, which is governed by group interactions. It is namely known that games governed by group interactions can yield a qualitatively different outcome from games that are governed by pairwise interactions \cite{perc_jrsi13}. The public goods game captures the essence of a social dilemma in sizable groups. Within each group, cooperators contribute $1$ to the public good while defectors contribute nothing. All contributions are then summed up and multiplied by the synergy factor $r>1$, which takes into account the added value of a cooperative group effort. This implies that if all players in the group choose the same strategy, they are better off cooperating than defecting. Subsequently, the accumulated goods are divided equally among all group members irrespective of their strategies to yield the payoff $\pi_x^g$ on network A and $\pi_{x^\prime}^g$ on network B. Note that defectors bear no costs when collecting identical benefits as cooperators. However, if nobody cooperates the synergy factor has not effect and the public goods are lost, hence resulting in a social dilemma. The public goods game is staged on a $L \times L$ square lattice with periodic boundary conditions, where players are arranged into overlapping groups of size $G=5$. Every player is thus surrounded by its $k=G-1$ nearest neighbors and is a member in $g=G$ different groups. As described, the payoff obtained in each group is $\pi_x^g$ and $\pi_{x^\prime}^g$, while the total amount each player $x$ receives from all the five groups is given by the sum $\pi_x = \sum_g \pi_x^g$ on network A and $\pi_{x^\prime} = \sum_g \pi_{x^\prime}^g$ on network B, which runs over all the $g=G$ different groups where $x$ is member. All the other details of the model are identical with those outlined for the prisoner's dilemma game.

Simulations were performed by means of a random sequential update, where each player on both networks had a chance to pass its strategy once on average during a Monte Carlo (MC) step. The linear system size was varied from $L=800$ to $6400$ in order to avoid finite size effects, and the equilibration required up to $10^6$ MC steps. Further simulation details are provided in the figure captions and the Results section.

\section{Results}
\label{results}
\begin{figure}
\centerline{\epsfig{file=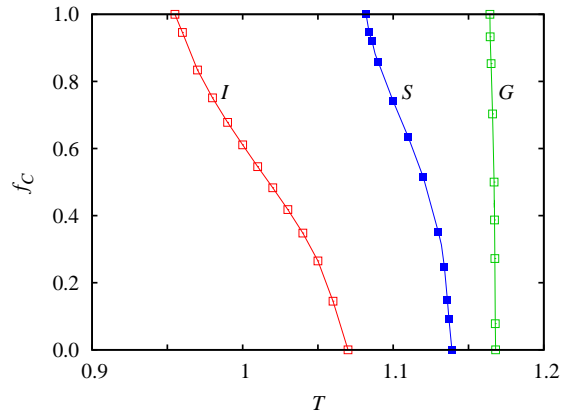,width=8cm}}
\caption{\label{pd} Information sharing promotes cooperation in the prisoner's dilemma game. The more information is shared, the higher the temptation to defect $T$ at which cooperators are able to survive. The critical $T$ at which cooperators are able to dominate completely increases as well, while the extent of the mixed $C+D$ phase shrinks. Depicted is the fraction of cooperators $f_C$ in dependence on $T$ for models $I$, $S$ and $G$, as denoted in the figure.}
\end{figure}

Throughout this section, we continuously compare the outcomes of the previously described models denoted as $I$, $S$ and $G$, which differ in to what extent information is shared between players on the two networks. Accordingly, in model $I$ information is not shared, and the evolution on both networks proceeds completely independently. In model $S$ information is shared only between single players, while in model $G$ the shared information is collected from a whole group. Model $G$ can be interpreted as an ``information cloud'' from another network helping the player to adjust the propensity to change strategy, as schematically depicted in Fig.~\ref{scheme}. As Fig.~\ref{pd} shows, the temptation to defect at which cooperators die out increases steadily with the amount of information that is shared. Model $I$ recovers the traditional spatial prisoner's dilemma game, while models $S$ and $G$ strongly favour the odds of cooperators, also by increasing the largest allowable temptation to defect at which cooperators die out completely. The mixed phase region sustaining a stable state of cooperators and defectors shrinks, which indicates that besides traditional network reciprocity additional, possibly more subtle, mechanisms are at work.

\begin{figure}
\centerline{\epsfig{file=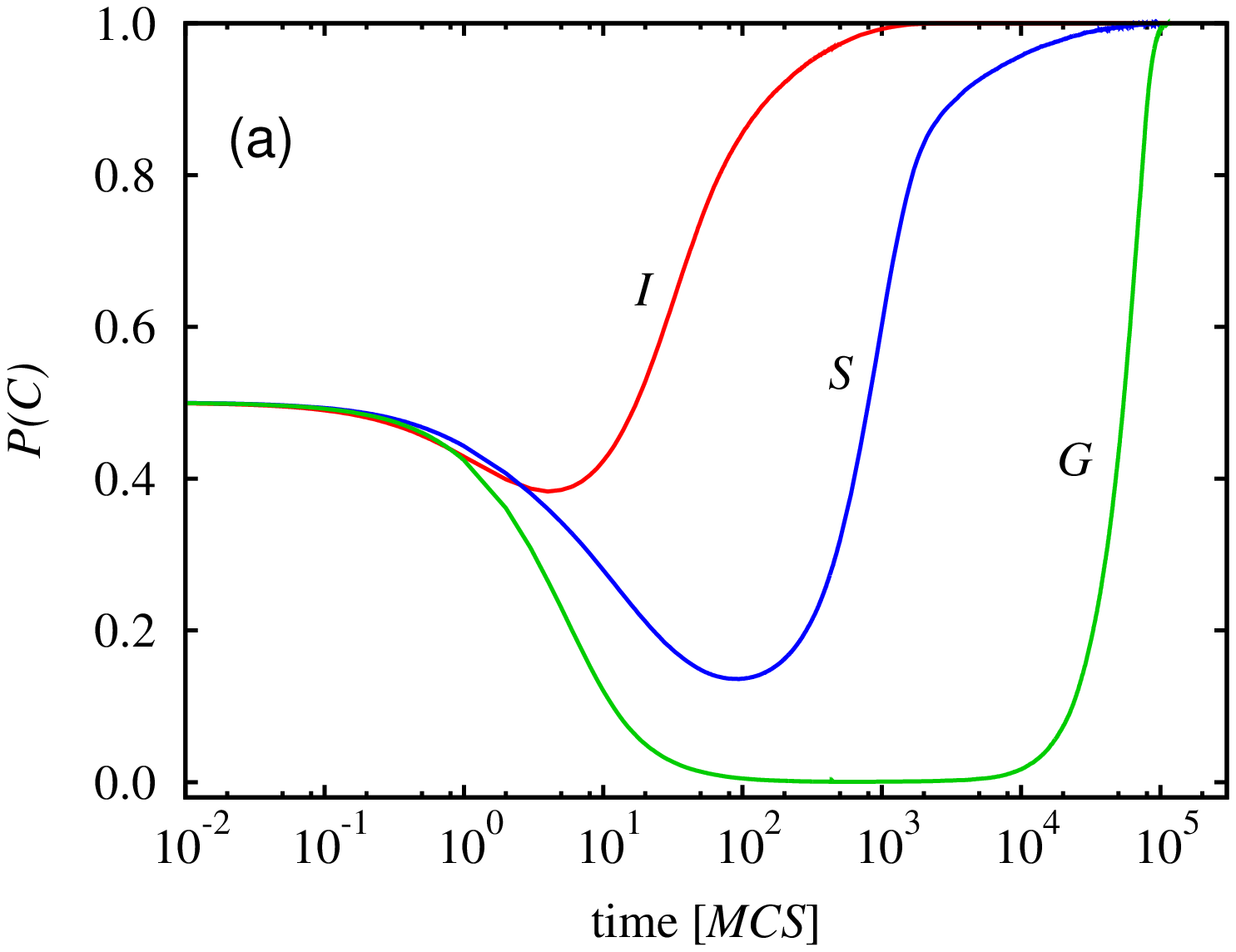,width=6cm}\epsfig{file=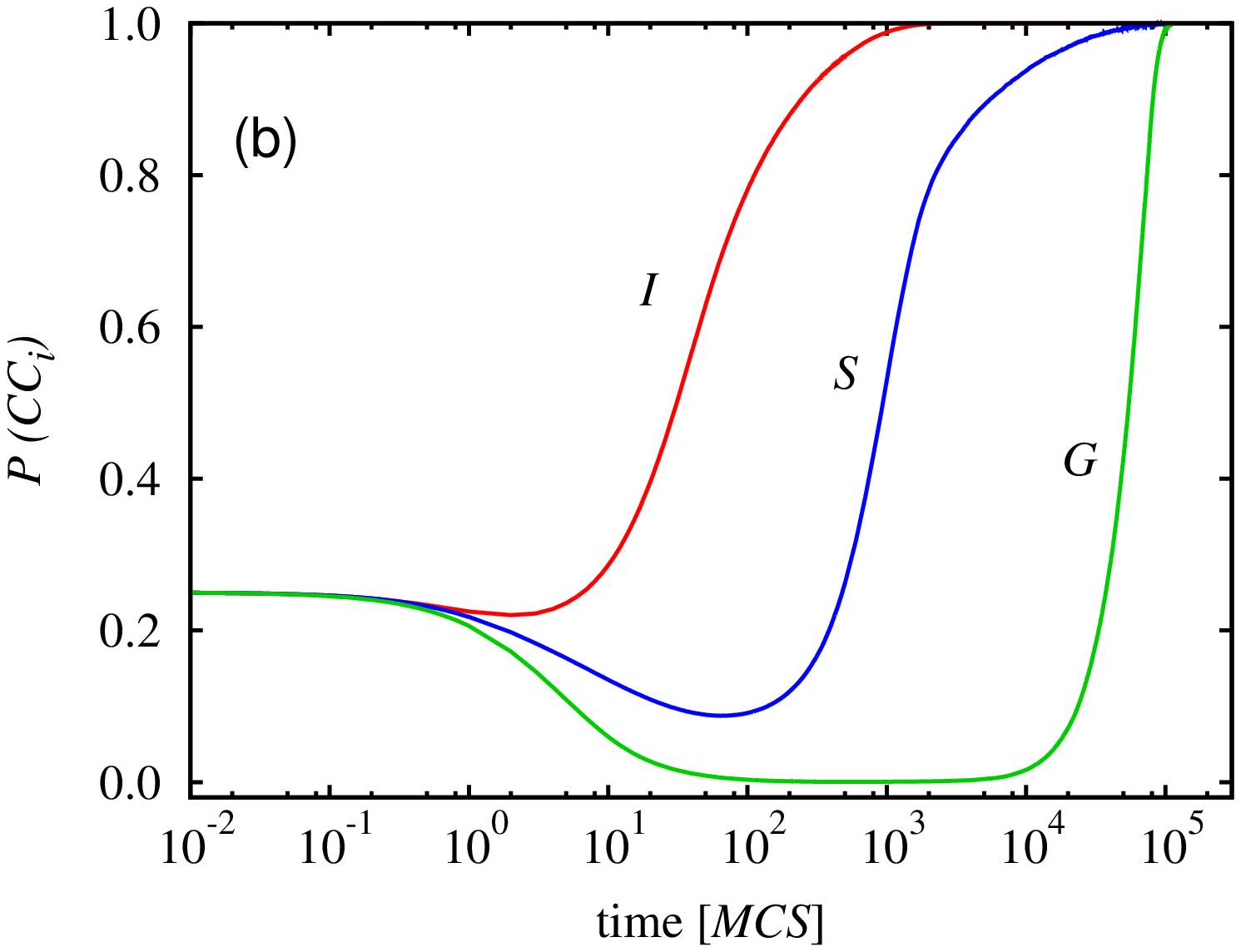,width=6cm}}
\centerline{\epsfig{file=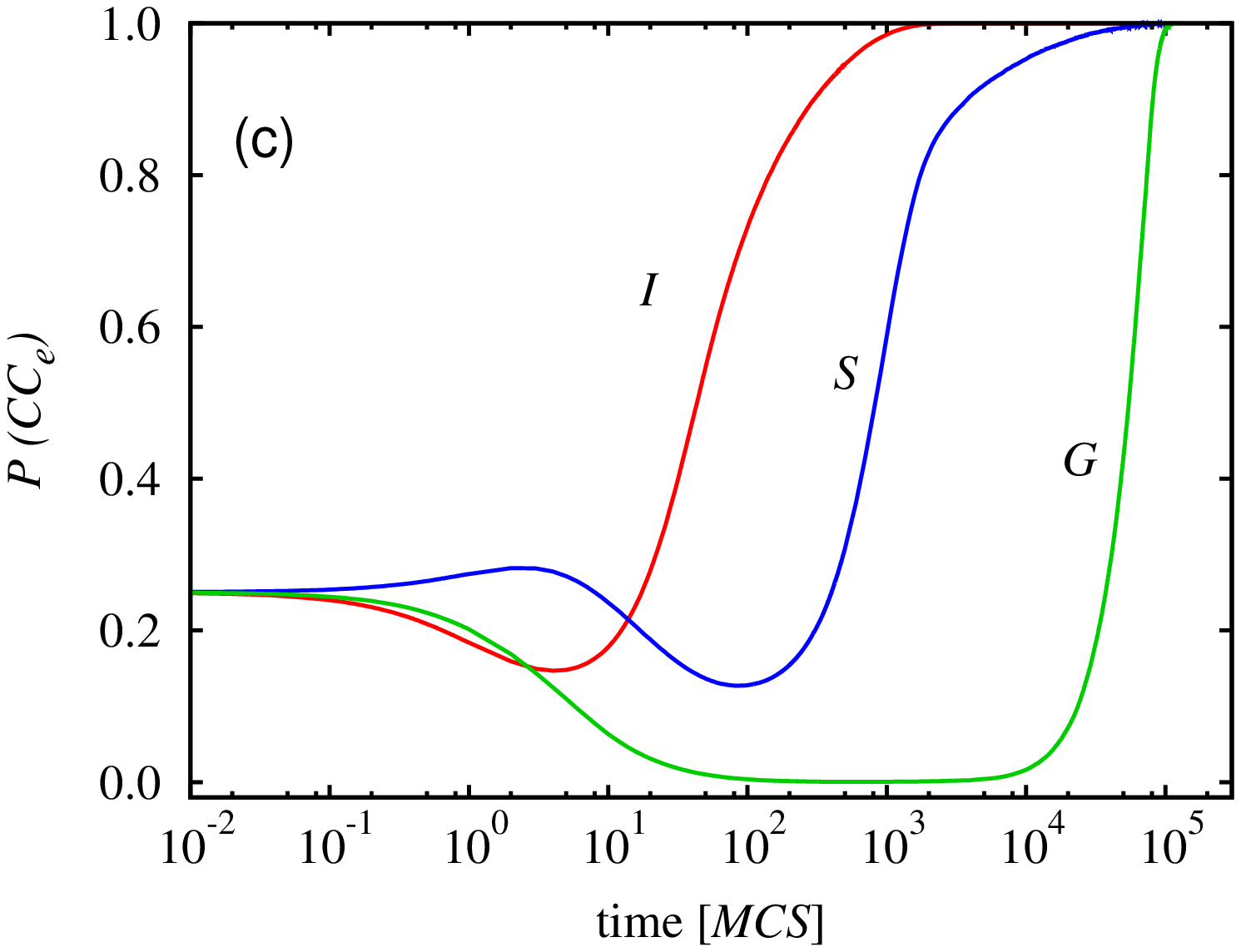,width=6cm}\epsfig{file=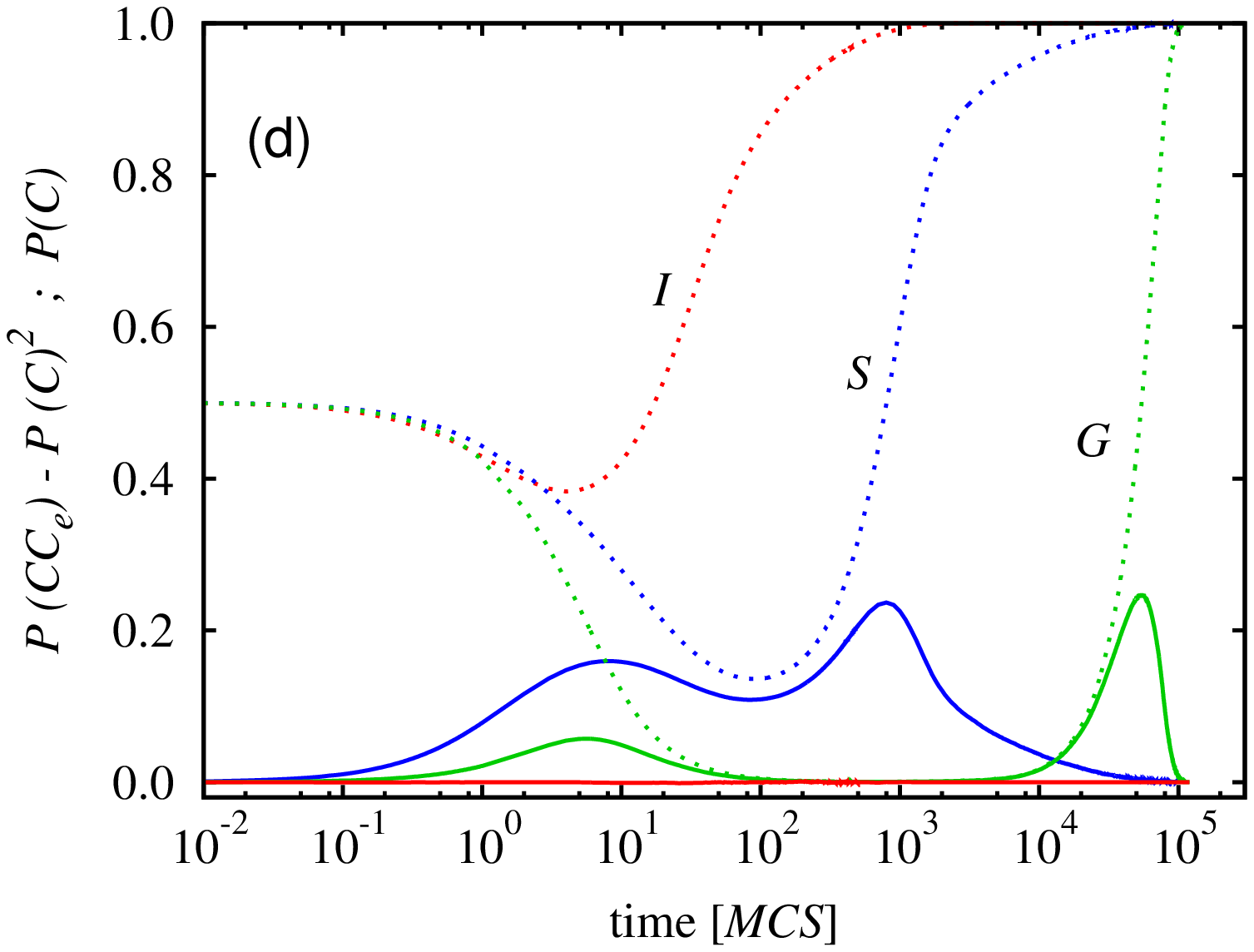,width=6cm}}
\caption{\label{evol} Time evolution of cooperation and cooperative pairs within and between networks reveals why information sharing promotes cooperation. Panel (a) shows the fraction of cooperators $P(C)$ in dependence on time, panel (b) shows the fraction of $C-C$ pairs within a network $P(CC_i)$, while panel (c) shows the fraction of $C-C$ pairs between the two networks $P(CC_e)$. Solid lines in panel (d) depict the excess correlation of cooperators between the two networks, determined as $P(CC_e)-P(C)^2$. For easier reference, panel (d) also features $P(C)$ as dotted lines. The values of $T$ are $0.95$, $1.08$ and $1.16$ for models $I$, $S$ and $G$ (as denoted on the figure), respectively. We have used up to $L=800$ system size and averaged the final outcome over $10$ independent runs to obtain accurate results.}
\end{figure}

To understand why cooperation is promoted in models $S$ and $G$ where information is shared, we compare the time evolution of different quantities when starting from a random initial state. To have an adequate comparison of the three different models, we choose values of $T$ that are close to the maximal values for which the system can still reach the full $C$ phase in the stationary state. The temptation is thus scaled so as to be effectively equal in all three models. In panel (a) of Fig.~\ref{evol}, we can observe the typical ``first down, later up'' trend of the fraction of cooperators, which is a trademark of network reciprocity. However, while the dip of $P(C)$ for model $I$ is small, models $S$ and $G$ exhibit a much stronger fall before cooperators are eventually able to recover. Obviously thus, when information is shared the mechanism that will eventually promote cooperation needs a while longer to start working effectively. And the delay is longer in the $G$ than in the $S$ model. We argue that the delay is due to the fact that more cooperators need to organize themselves when information is shared. Although utilities are not affected by players from different networks, the interdependence introduced by information sharing nevertheless imposes bonds that require coordination. Since in model $G$ more players are involved than in model $S$, it also takes more time. As the green line in Fig.~\ref{evol}(a) shows, cooperators almost die out before ultimately rising to dominance on both networks. This feature also involves that we must use a sufficiently large system size to obtain reliable results for model $G$, especially in the vicinity of phase transition points.

Since the fraction of cooperators fails to convey information about correlations between the strategies, we present in panel (b) of Fig.~\ref{evol} how the fraction of $C-C$ pairs within a network $P(CC_i)$ varies with time. For model $I$ the curve virtually does not decline before the rise, indicating that if $C-C$ pairs are present, they can immediately establish the necessary conditions to spread based on traditional network reciprocity. For $S$ and even more so for the $G$ model, however, the sole vicinity of cooperators within a network does not ensure the necessary conditions for spreading. In fact, due to significantly higher temptation values, network reciprocity alone would be unfit to prevent the extinction of cooperators. Accordingly, an additional mechanism must emerge for the downward trend to reverse, and as expected, it comes from the information shared between the two networks.

In order to reveal this, panel (c) of Fig.~\ref{evol} shows how the fraction of $C-C$ pairs between the two networks $P(CC_e)$ changes during the evolution. However, it is conspicuous that for model $S$ this curve shows a slight temporary maximum after only two or three $MC$ steps. The early maximum is due to randomly established $C-C$ pairs, but their mutual support is fragile and cannot be maintained in the absence of local clustering within a network. For model $G$ this effect is less pronounced because it is very unlikely that the whole group will be in the pure $C$ state initially. The real correlation between the two networks is shown in panel (d) of Fig.~\ref{evol}, where we compensate for the fact that $P(CC_e)$ can be high even if there is no information transfer between the two networks. To avoid this, we subtract $P(C)^2$ from $P(CC_e)$, which then yields the probability to find excess $C-C$ pairs between the two networks. Expectedly, this quantity is always zero in model $I$, as there the promotion of cooperation is due solely to traditional network reciprocity. For the models $S$ and $G$, however, the emergence of excess $C-C$ pairs between the two networks is crucial, as it enables the rise of cooperators from their initial decline. To enable a direct comparison, we depict in panel (c) also the fraction of cooperators as dotted lines. As we have already noted, there are some correlations between cooperators at the very beginning of the evolutionary process, especially in model $S$, but this additional effect that emerges due to information sharing can be really powerful only if it goes hand in hand with the clustering of cooperators within a network. By comparing the curves with those presented in panel (b), we arrive at the conclusion that only clustering \textit{and} excess correlations due to information sharing can overcome defectors at temptations to defect that exceed those that can be offset by network reciprocity alone.

\begin{figure}
\centerline{\epsfig{file=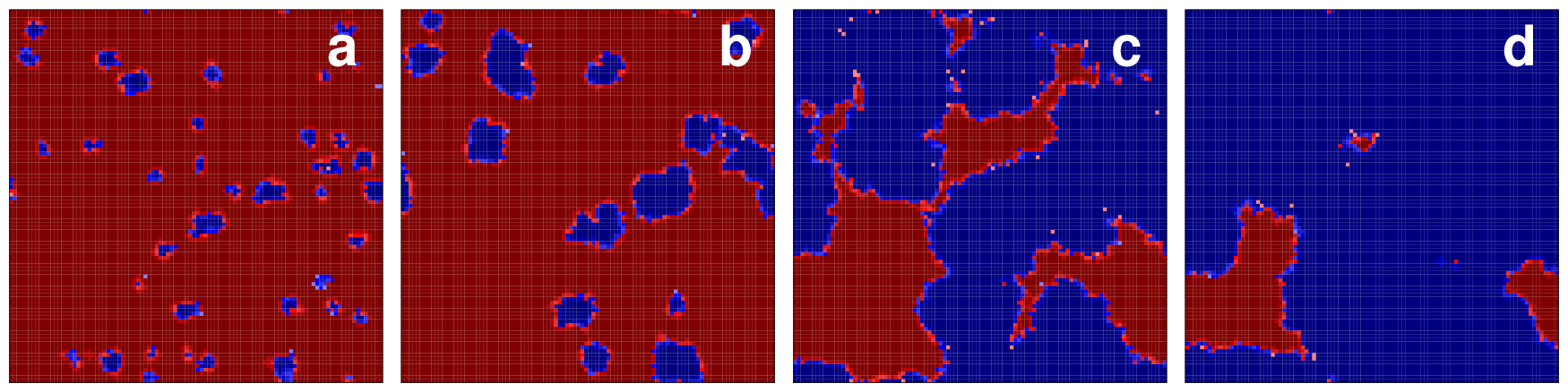,width=15cm}}
\centerline{\epsfig{file=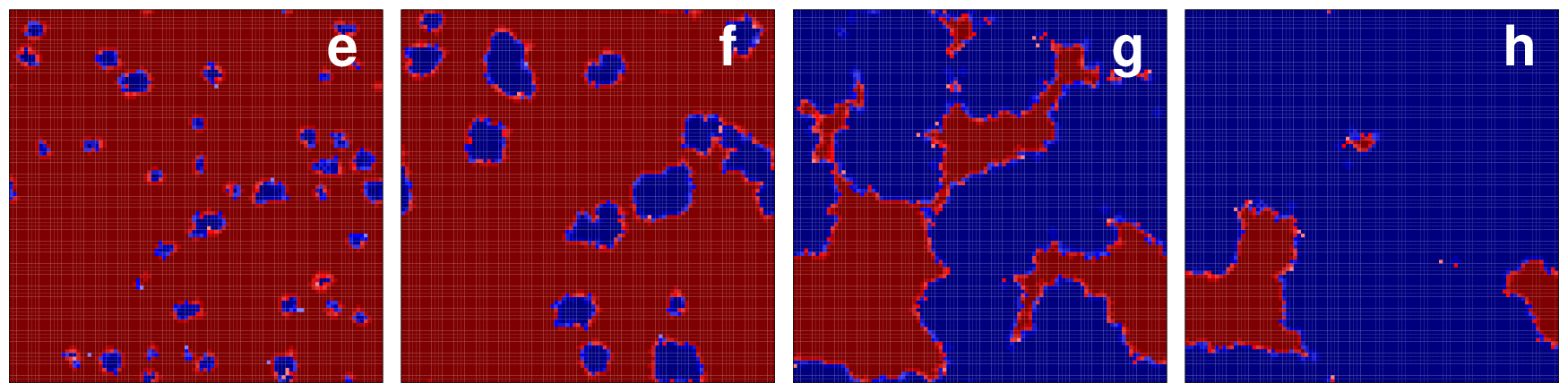,width=15cm}}
\caption{\label{G} Characteristic snapshots reveal the spontaneous emergence of correlated evolution that is due to information sharing. Presented are snapshots of the upper (a-d) and lower (e-h) network for model $G$ as obtained for $T=1.08$ after $50$, $250$, $700$ and $1000$ MCS from left to right. Defectors are denoted red and cooperators are denoted blue. To visualize different values of the scaling factors $w_x$ that modify the propensity of each player to change strategy, we use different shades of red and blue, where darker colour marks smaller and brighter colour marks larger values of $w_x$. The random initial state and the final pure $C$ phase are not shown.}
\end{figure}

\begin{figure}
\centerline{\epsfig{file=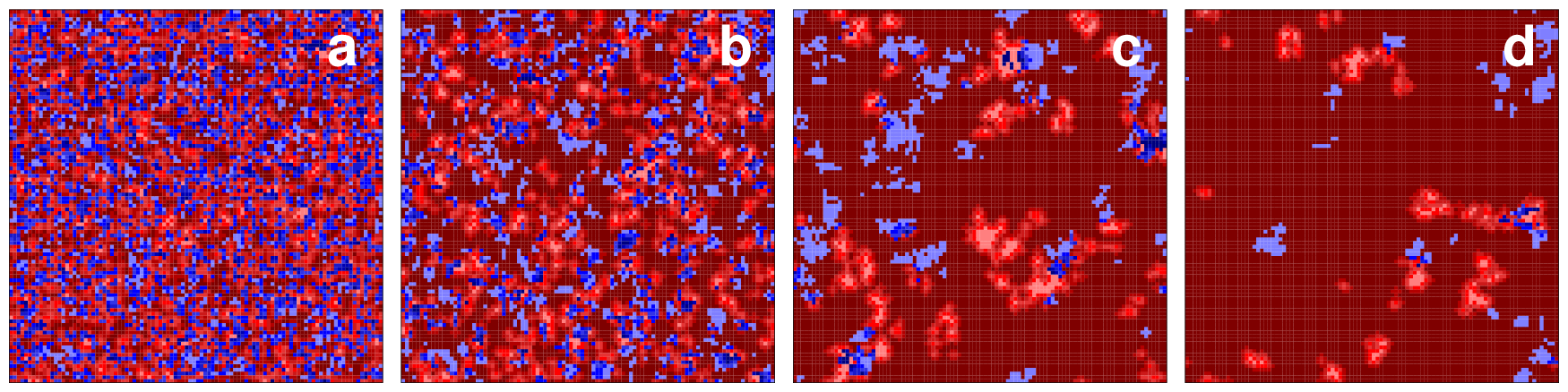,width=15cm}}
\centerline{\epsfig{file=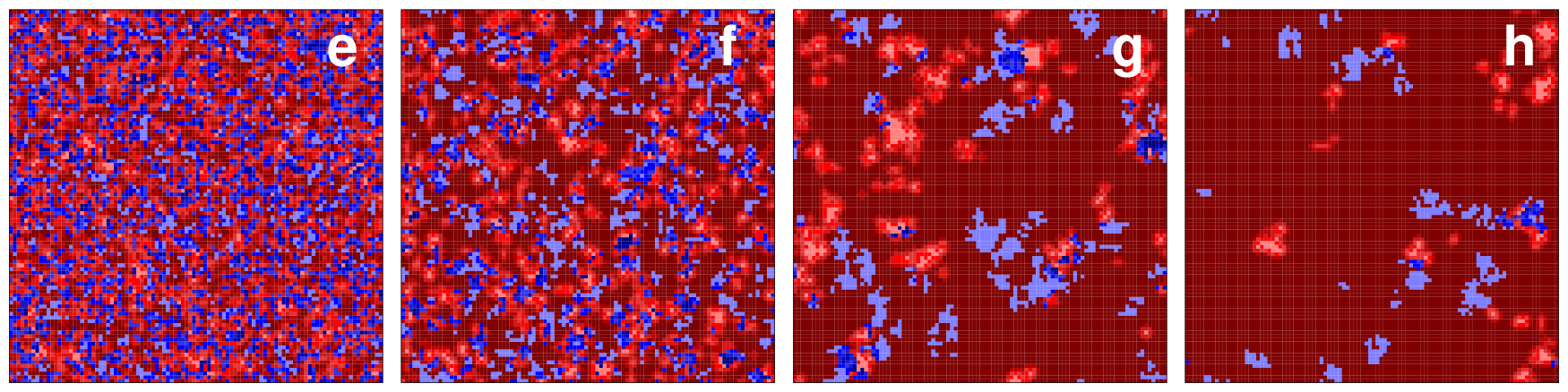,width=15cm}}
\caption{\label{I} In the absence of information sharing evolution on the two networks proceeds uncorrelated. Presented are snapshots of the upper (a-d) and lower (e-h) network for model $I$ as obtained for $T=1.08$ after $2$, $10$, $100$ and $300$ MCS from left to right. The applied colour scheme is the same as for Fig.~\ref{G}. Although $w_x=1$ for all players, we use the same shading to highlight the uncorrelated evolution. The random initial state and the final pure $D$ phase are not shown.}
\end{figure}

The emergence of correlations between the two networks can be visualized by examining characteristic snapshots as they evolve on them, as presented in Fig.~\ref{G} for model $G$. These snapshots clearly emphasize that the extra correlations between the two networks, in association with the clustering within a network, warrant a powerful support for cooperators to spread. Indeed, the correlation in model $G$ is conspicuous, as we can observe almost an identical evolution taking place on both networks, despite the fact that only the information about strategies is shared solely in a way that affects the propensity of players to change their strategy. Neither payoffs nor actual strategies are transmitted between the two networks. For comparison, we have plotted the same series of snapshots as obtained with model $I$ for the same set of parameters. In this case, as Fig.~\ref{I} shows, there is no correlation emerging between the two networks. Note that here light blue domains reflect the absence of cooperators in the same place on the other network, while light red signals the same for defectors. Darker shades of red in the final stages are not due to correlation, but simply because defectors are very widespread [this also further supports the subtraction of $P(C)^2$ to quantify correlations between the two networks in Fig.~\ref{evol}(d)]. While cooperators do try to aggregate into compact clusters on both isolated networks, the network reciprocity alone is unable to prevent their extinction at such a high temptation to defect.

\begin{figure}
\centerline{\epsfig{file=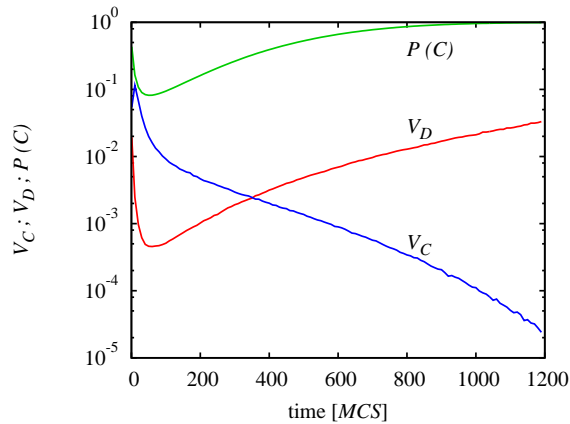,width=8cm}}
\caption{\label{vulnerable} Vulnerable players that do not get affirmative information from the partner network decide the fate of evolution. While cooperators advance very slowly, they also do so compactly and coherently on both networks. They therefore reinforce their strategies when sharing information between the networks. Defectors, on the other hand, proceed fast and with disregard to what their partners on the other network are doing. Ultimately this is their demise. Depicted is the fraction of vulnerable (those having $w_x=w_{\max}=1$) cooperators $V_C$ and defectors $V_D$, as well as the overall fraction of cooperators $P(C)$, as obtained with model $G$ for $T=1.08$ at $L=3200$ system size.}
\end{figure}

We have thus established that the correlated pattern formation, evoked by the information sharing between the two networks, plays a fundamental role in ensuring promotion of cooperation past the boundaries imposed by traditional network reciprocity. However, given that information sharing is implemented without any assumptions with regards to content, i.e., regardless of the strategy of the player or the group of players that transmit the information, one could still be curious as to why the procedure favours cooperators but not defectors. After all, given the correlated evolutionary process on both networks, defectors could inform each other about the success of their strategy too. Moreover, $D-D$ pairs forming between the two networks also protect each other by reducing their propensity to adopt the occasionally more successful cooperative strategy. To clarify the biased consequence of strategy neutral information sharing, we monitor the relative fraction of the so-called vulnerable players in the $G$ model. We designate as vulnerable every player who has $w_x=1$, because there is no support coming from the other network in terms of reinforcing the player in the strategy it currently holds. Accordingly, such players are most likely to change their strategy, and are thus termed vulnerable. In Fig.~\ref{vulnerable}, we present the fraction of vulnerable cooperators $V_C$ and defectors $V_D$ over time, and contrast this with the overall fraction of cooperators $P(C)$. It can be observed that $V_C$ is really high, i.e., $V_C \approx P(C)$, during the early stages of the game. This is because cooperators are unable to form clusters at the beginning of the evolutionary process, which is a fundamental condition to establish an efficient support between the two networks. At the same time, the ratio for vulnerable defectors is extremely small because initially defectors spread very successfully on both networks, making full use of the recommendation power that is at their disposal due to information sharing. Yet shortly thereafter, the greediness of the defectors starts taking its toll. More precisely, defectors on an isolated network may proceed with their exploitation easily because of the high $T$ value. However, the ``cloud'' of cooperators supporting their spreading in the other network cannot follow such an aggressive invasion, and hence defectors become vulnerable in the absence of direct supporters in the other network. At the same time, cooperators employ a more careful strategy, which makes them invade slowly but compactly. As snapshots in Fig.~\ref{G} demonstrate, the borders of cooperative domains in both networks move almost simultaneously, and hence cooperators never leave related players in the other network ``unprotected''. Note that along the interfaces in Fig.~\ref{G} there are more bright red players than there are bright blue players. As the logarithmic vertical axis of Fig.~\ref{vulnerable} highlights, the ratio of vulnerable defectors is very small when it changes tendency, yet still the rise of $V_D$ is followed immediately by an overall increase of the cooperation level.

\begin{figure}
\centerline{\epsfig{file=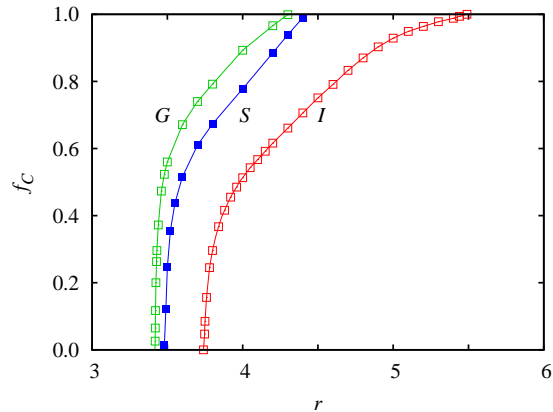,width=8cm}}
\caption{\label{pgg} Information sharing promotes cooperation also in games that are governed by group interactions. The more information is shared, the lower the synergy factor $r$ at which cooperators are able to survive. The critical $r$ at which cooperators are able to dominate completely decreases as well. Compared to the effect reported for the prisoner's dilemma game in Fig.~\ref{pd}, however, for the public goods game the impact of information sharing is weaker because group interactions do not allow as large spatial fluctuations that would result in abundant vulnerable defectors. Depicted is the fraction of cooperators $f_C$ in dependence on $r$ for models $I$, $S$ and $G$, as denoted on the figure.}
\end{figure}

Since the prisoner's dilemma game is governed by pairwise interactions, it is finally of interest to clarify the role of information sharing also in games that are governed by group interactions. In the Model section, we have introduced the public goods game as a classical example of a game that is governed by group interactions. As recently reviewed in \cite{perc_jrsi13}, games governed by group interactions can yield very different evolutionary outcomes on structured populations than games that are governed by pairwise interactions. Notably, due to the participation in the same groups, even players that are not physically connected by links act as if they were, which in turn introduces qualitatively different limits in terms of the impact of uncertainty by strategy adoptions \cite{szolnoki_pre09c}, as well as in terms of spatial fluctuations which are effectively  averaged out \cite{perc_njp11}. As results presented in Fig.~\ref{pgg} demonstrate, information sharing does promote the evolution of cooperation also in the public goods game. Yet the impact is rather modest, especially when going from model $S$ (information is shared only between single players) to model $G$ (shared information is collected from a whole group). This is a direct consequence of multi-point interactions, which preclude large spatial fluctuations and thus hinder the emergence of vulnerable defectors to the same extent as reported above for the prisoner's dilemma game. Vulnerable defectors do of course emerge, but the interfaces are much more blurred due to group interactions, hence allowing defectors to receive more support from the other network than they do by pairwise interactions. Nevertheless, the excursion to games governed by group interactions does confirm that sharing information is certainly not harmful, and thus can be fully recommended in support of prosocial behaviour.

\section{Discussion}
\label{discussion}
We have introduced information sharing to evolutionary games. Players residing on different, not physically connected networks, are allowed to exchange information in the form of strategies they are adopting, and this information regulates the propensity of the receiver to change its strategy. We have shown that information sharing promotes prosocial behaviour past the limits of traditional network reciprocity. The effect is due to the synergy between network reciprocity and the spontaneous emergence of correlated behaviour on the two networks. If information is shared only between single players, the maximally allowable temptation to defect that still allows cooperators to survive increases markedly. Further improving the odds of cooperation is if the information is shared not just between single players but rather if it stems from groups. Importantly, we have demonstrated that no additional assumptions are needed with regards to the information that is shared for the newly identified mechanism to take effect. Defectors are just as free to inform others about their strategy as cooperators. Ultimately, it is the excessive greediness and disregard for the neighbours that dooms defectors. While during the early stages of the game defectors may exploit the population more effectively than in the absence of information sharing, the trend reverses sharply after both the network reciprocity and the correlated behaviour between the two networks set in. The way of cooperation is slow but steady, and as such it is designed so as to make optimal long run use of information sharing.

Although information sharing is less effective for the promotion of cooperation in games that are governed by group interactions, we have shown that it nevertheless does help. This bodes well for the general applicability of information sharing as a means to promote prosocial behaviour. However, one might wonder what happens if the size of groups acting as information sources increases. Large groups effectively act as mean-field territory, and as such they seem unable to advise relevantly on the subject of strategy change. The information stemming from large conglomerates is often diluted to a degree that it cannot serve a particular purpose. While crowdsourcing and the so-called ``wisdom of the crowd'' effect \cite{surowiecki_04} have recently received ample attention \cite{rauhut_jmp10,golub_aejm10,lorenz_pnas11}, also in the realm of evolutionary games \cite{szolnoki_srep12}, in the context studied here they cannot be advocated as viable choice. Sharing information locally is more akin to the consideration of reputation \cite{fehr_n04,fu_pre08b,pfeiffer_jrsi12}, where players decide how to proceed based on their experience from the past, and in doing so, they are able to deduce their expected payoff in the future. Cooperators prefer colonization, and they do well in maintaining their good reputation. Receiving information that a group is adopting cooperation is thus inherently different from the information that a group is adopting defection. In the latter case, although it may suggest that to defect is a good idea, it quickly turns out that it is not, as the shadow of the past is quick to catch up in the absence of new cooperators that one could exploit. It is thus reassuring to discover that the nature of prosocial behaviour in social dilemmas is such that it naturally makes optimal use of information stemming from other sources, if only it is made available.

\ack
This research was supported by the Hungarian National Research Fund (Grant K-101490) and the Slovenian Research Agency (Grant J1-4055).

\section*{References}
\providecommand{\newblock}{}

\end{document}